\documentclass[a4paper]{article}
\usepackage{geometry}
\usepackage{amsmath}
\usepackage{amssymb}
\usepackage{amsfonts}
\usepackage{bm}

\usepackage{graphicx,import}
\usepackage{varwidth}
\usepackage{calc}
\usepackage[export]{adjustbox}
\usepackage{booktabs}
\usepackage[ruled,vlined]{algorithm2e}% for pseudocode

\usepackage{color,soul}
\usepackage{hyperref}
\usepackage{enumitem}
\usepackage{pst-all}
\usepackage{layouts}
\usepackage{scalerel}
\usepackage{setspace}
\usepackage{authblk}
\usepackage{nomencl}
\usepackage{fancyhdr}
\usepackage{tikz}
\usetikzlibrary{patterns,shapes.arrows}
\usetikzlibrary{arrows.meta}
\usepackage{pgfplots}

\usepackage[
backend=biber,
style=numeric,
sorting=ynt
]{biblatex}

\usepgfplotslibrary{groupplots}
\pgfplotsset{compat=newest}
\usepackage[intlimits]{esint}

\usepackage{nomencl}
\makenomenclature

\usepackage{etoolbox}
\renewcommand\nomgroup[1]{%
  \item[%\bfseries
  \ifstrequal{#1}{R}{\textit{Roman letters}}{%
  \ifstrequal{#1}{G}{\textit{Greek letters}}{%
  \ifstrequal{#1}{D}{\textit{Dimensionless groups}}{%
  \ifstrequal{#1}{S}{\textit{Superscripts and subscripts}}{%
  \ifstrequal{#1}{A}{\textit{Abbreviations}}{}}}}}%
]}
\begin{document}
\pagestyle{fancy}
\fancyhead{}
\fancyfoot{}
%\fancyfoot[L]{Jaeschke}
\fancyfoot[C]{\thepage}
\fancyfoot[R]{Jaeschke et al.}

\title{\textbf{Effect of Additively Manufactured Wall Lattice Structures on Flashback Limits in a Hydrogen Jet Flame Combustor}}

\author[1]{Alexander\ Jaeschke\thanks{a.jaeschke@tu-berlin.de}}
\author[2]{Thomas Ludwig Kaiser}
\author[3]{Lukas Melzig}
\author[3]{Michael F. Zaeh}
\author[2]{Kilian Oberleithner}
\author[1]{Christian Oliver Paschereit}

\affil[1]{Chair of Fluid Dynamics – Technische Universität Berlin, Germany}
\affil[2]{Laboratory for Fluid Instabilities and Dynamics – Technische Universität Berlin, Germany}
\affil[3]{Institute for Machine Tools and Industrial Management (\textit{iwb}) – Technical University of Munich, Germany}
\date{}
\maketitle

\providecommand{\keywords}[1]
{
  \small	
  \textbf{\textit{Keywords---}} #1
}

\keywords{gas turbine, hydrogen combustion, jet flame, flashback, PIV, additive manufacturing, PBF-LB/M}

%%%%%  ABSTRACT  %%%%%%%%%%%%%%%%%%%%%%%%%%%%%%%%%%%%%%%%%%%%%%%%%%%
\doublespacing
\onecolumn
\begin{abstract}
This study investigated how additively manufactured nozzles with body-centered cubic lattice structures reduce the flame flashback propensity in a hydrogen jet flame burner. Five different configurations of a jet flame combustor were investigated, with a focus on mixing duct walls incorporating porous media. The nozzles were manufactured by the powder bed fusion of metals using a laser beam process. 
The lattice parameters were varied by the volume fraction and the strut diameter. For the experiments, pure hydrogen was used as fuel under atmospheric conditions at various equivalence ratios and Reynolds numbers of 9,000 -- 12,000.
Flow field measurements, flame imaging, and spectral proper orthogonal decomposition of the flame dynamics were employed to identify possible transition mechanisms from a stable operation to flashback. The flow fields and the flame shapes showed only minor effects from wall modifications, preserving general flow characteristics across configurations. The flow dynamics in the combustion chamber were dominated by large-scale coherent structures in the shear layer, specifically Kelvin-Helmholtz instabilities. 
The results demonstrated that the nozzle with the coarsest porous wall structure significantly improved the flashback resistance compared to a nozzle with a solid wall.
It is concluded that the primary mitigation mechanism was a cooling effect by unburnt mixture flowing through the porous media. 
The findings confirmed that the integration of lattice structures through additive manufacturing provides a viable strategy for hydrogen flashback mitigation by manipulating the coupled interaction between the flame and the thermal conditions of the wall.

\end{abstract}

\nomenclature[R]{\(\omega\)}{Frequency [Hz]}
\nomenclature[R]{\(\text{S}_{\text{L}}\)}{Laminar flame speed (unstretched) [m/s]}
\nomenclature[R]{\(u, v\)}{Velocity components [m/s]}

\nomenclature[G]{\(\nu\)}{Kinematic viscosity [m$^2$ s$^{-1}$]}
\nomenclature[G]{\(\kappa\)}{Thermal conductivity [W m$^{-1}$ K$^{-1}$]}
\nomenclature[G]{\(\lambda_\omega^{(j)}, \widehat{\psi}_\omega^{(j)}\)}{SPOD mode and eigenvalue of mode $j$}

\nomenclature[D]{\(\phi\)}{Equivalence ratio}
\nomenclature[D]{\(\text{Re}\)}{Reynolds number, $ud/\nu$}
\nomenclature[D]{\(\text{St}\)}{Strouhal number, $fd/u$}

% \EntryHeading{Superscripts and subscripts}
\nomenclature[S]{\(0\)}{Initial condition}
\nomenclature[S]{\(\text{ad}\)}{Adiabatic}
\nomenclature[S]{\(\text{B}\)}{Block}
\nomenclature[S]{\(\text{f}\)}{Fraction}
\nomenclature[S]{\(\text{L}\)}{Layer}
\nomenclature[S]{\(\text{q}\)}{Quenching}
\nomenclature[S]{\(\text{rms}\)}{Root mean square}

\nomenclature[A]{\(\text{AM}\)}{Additive manufacturing}
\nomenclature[A]{\(\text{BCC}\)}{Body-centered cubic}
\nomenclature[A]{\(\text{FB}\)}{Flashback}
\nomenclature[A]{\(\text{OH*}\)}{Excited hydroxyl radical}
\nomenclature[A]{\(\text{PIV}\)}{Particle image velocimetry}
\nomenclature[A]{\(\text{PBF-LB/M}\)}{Powder bed fusion of metals using a laser beam}
\nomenclature[A]{\(\text{PSD}\)}{Power spectral density}
\nomenclature[A]{\(\text{SPOD}\)}{Spectral proper orthogonal decomposition}

\printnomenclature

%%%%%%%%%  BODY OF PAPER %%%%%%%%%%%%%%%%%%%%%%%%%%%%%%%%%

\section*{Introduction}
Hydrogen presents both significant promise in carbon dioxide reduction and substantial engineering challenges for the combustion in gas turbines. As renewable energy sources increasingly dominate the energy sector to replace fossil fuels, unpredictable fluctuations and seasonal availability variations emerge. Gas turbines offer an efficient way to enhance the flexibility in renewable energy-based power systems. To maintain low carbon emissions, hydrogen is the preferred alternative to replace natural gas as fuel. However, the high reactivity and flame temperature of hydrogen allow only limited quantities in fuel mixtures before flame flashback (FB) occurs in swirl-stabilized industrial gas turbines~\cite{Ciani2019,magnusson:2020,cosic:2022}.

Increasing the axial velocity prevents the flame propagation upstream into regions which are not designed to withstand flame anchoring. Jet-stabilized flame concepts therefore represent a promising solution for hydrogen combustion \cite{noble:2021,jaeschke:2023,hermeth:2024, zhou:2024}. Fully exploiting the potential of hydrogen requires addressing flame flashback, nitrogen oxide (NO$_{\text{x}}$) emissions, and dynamic instability. Additive manufacturing (AM) has the potential to make a substantial contribution to overcome these challenges due to the significant design freedom it offers \cite{hermeth:2024,jaeschke:2025, ax:2025}.

For turbulent hydrogen jet flames, boundary layer FB is the dominant mechanism \cite{kalantari:2015, Eichler:2011}. The interaction of the inlet velocity, the turbulent flame speed, and the flame quenching distance can cause the flame to propagate upstream along the wall when the boundary layer velocity falls below the burning velocity in proximity to the wall, known as the critical gradient concept \cite{Elbe:1945, kalantari:2015}. Recent studies draw a more complex picture of the process revealing that velocity fluctuations initiate the onset of FB \cite{porath:2025, park:2025} and that backflow regions form ahead of the FB propagating into the mixing tube \cite{Eichler:2011, gruber2021}. Thereby, the elevated burning velocity of hydrogen stems not solely from its higher laminar flame speed but also from thermal-diffusive instability \cite{berger:2022}, which drives the characteristic conical shape of the flame across a wide range of operating conditions.

Studies by Duan et al.~\cite{Duan:2014} and Baumgartner et al.~\cite{Baumgartner:2015} reported a strong influence of the burner material, the flame confinement, and the burner tip temperature on the FB propensity. They proposed burner rim cooling to prevent FB but also suggested low-conductivity injector material or coatings for an improved thermal stability. Based on these results and comprehensive FB measurements a low-order model was proposed by Kalantari et al.~\cite{kalantari:2017} but it cannot reliably predict FB limits with parameters far from the initial conditions. Significant improvements were done by Hoferichter et al.~\cite{hoferichter2017} but the transferability to other configurations is still a limiting factor.   

Park et al.~\cite{park:2025} reevaluated in their recent study the influence of the wall heat flux and its relationship to FB dynamics for an unconfined jet flame burner. They confirmed the findings of Duan et al.~\cite{Duan:2014} with a more detailed look at the quantification of the wall heat flux and the resulting quenching distance for different burner cooling strategies. They stated that a comparison with data from the literature is only possible on a qualitative basis revealing the critical influence of trivial adjustments and setup details in FB studies.

Strategies for the detection of FB in a single-jet hydrogen burner were reported by zur Nedden et al.~\cite{zurnedden:2025} aiming for a reliable method for early FB detection with promising results based on measurements of the material resistance at the mixing tube rim.

Ax et al.~\cite{ax:2025} reported on a single-nozzle configuration and its feasibility of operating with 100~\% hydrogen at gas turbine relevant conditions. They stated that the transferability of the reported FB limits to full-scale multi-nozzle gas turbine burner might be difficult because of the influence of adjacent flames. Key aspects identified in the study were the surface roughness of the mixing duct and a local increase in the heat transfer on the nozzle rim.

Investigations on multi-jet burners confirmed the potential of jet-stabilized combustion systems composed of individual single jets. FB limits, NO$_{\text{x}}$ emissions, and dynamic stability were investigated for various setups reporting promising results for hydrogen combustion \cite{jaeschke:2023,beuth2023, kang:2021, jin:2021}.
Full-engine tests of a piloted multi-nozzle combustor under full-load conditions were recently published by Hermeth et al.~\cite{hermeth:2024}.
The reported fuel-flexible system was capable of operating with 100~\% hydrogen showing promising results on the flame stability, emissions, and the FB risk. However, the tests were limited by head-end temperature constraints imposed by the FB risk. Progress in reaching full-load conditions with 100~\% hydrogen were also reported by \cite{ge:2025}.

Despite intense research, many questions on the onset of FB are still unclear. Recent investigations by von Saldern et al.~\cite{vonsaldern2025} identified low-frequency streaky structures in a lean hydrogen flame raising the question of how jet dynamics impact the FB behavior.

The present study investigated additively manufactured nozzle designs for hydrogen flame FB mitigation in jet flame burners, with a focus on mixing duct walls incorporating porous media. The central question was to what extent the porous wall sections increase the FB resistance and what the dominant underlying mechanism is.
Nozzle variants were compared under both stable and unstable operating conditions, and were evaluated against known FB indicators, such as the burner rim temperature. Flow field measurements and flame imaging were used to characterize the influence of porous media on the combustion process, while flow dynamics were analyzed to identify possible transition mechanisms from a stable operation to FB.

%%%%%%%%%%%%%%%%%%%%%%%%%%%%%%%%%%%%%%%%%%%%%%%%%%%%%%%%%%%

\section{Materials and methods}\label{sec:exp_setup}
\subsection{Experimental setup}\label{sec:nozzles}

\begin{figure}
    \includegraphics{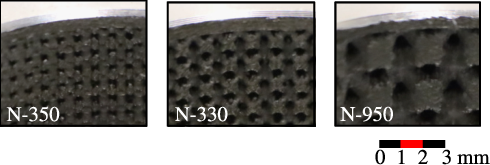}
    \caption{Close-ups of the porous walls with the solid plate at the nozzle outlet visible in the upper part of the images}
    \label{fig:mesh}
\end{figure}

In this study, five different configurations of jet flame burners were investigated by modifying the upper mixing duct section, called nozzle. 
The AM nozzles were made of Inconel~718 and comprised three different variants with a porous wall section, shown in Fig.~\ref{fig:mesh}, and one solid variant without porous media. 
The porous section had an axial length of 40~mm and a radial thickness of 4~mm. All AM porous nozzles had a solid outer wall in the radial direction.
In the axial direction, the porous section ended with a solid plate 1~mm thick at the nozzle outlet, which was flush with the burner plate and covered the porous area of the combustion chamber. Therefore, the entire gas mixture passed through the nozzle outlet into the combustion chamber.
For a comparison with conventionally manufactured nozzles and data from the literature, one nozzle, geometrically identical to the solid AM nozzle, was machined from 304 stainless steel (1.4301).

The porosity in the wall areas was realized through the integration of body-centered cubic (BCC) lattice structures into the nozzles. These lattice structures were characterized by their volume fraction $V_{\text{f}}$, their strut diameter $d_{\text{strut}}$, their surface area $A_{\text{cell}}$, and the resulting pore size $d_{\text{pore}}$, which is the free spacing between the lattice struts. The volume fraction is defined as the proportion of solid material within the total volume of a lattice structure. 
The surface area ratio, relative to a plain wall element of identical outer dimensions, is given by a constant of $A_{\text{cell}}/A_{\text{solid}}=4$. 
The naming of each nozzle in this study is based on its strut diameter and volume fraction, for example N-950 with $d_{\text{strut}}=0.9$~mm and $V_{\text{f}}=50$~\%. An overview of the dimensions of the lattice structures is given in Tab.~\ref{tab:lattice_structures}.

\begin{table}[!ht]
\caption[Table]{Overview of the dimensions of the lattice structures\label{tab:lattice_structures}}
\centering{%
\begin{tabular}{cccccc}
\toprule
\toprule
 &  $V_{\text{f}}$ & $d_{\text{strut}}$ & $d_{\text{pore}}$ & $A_{\text{cell}}$ & $A_{\text{cell}}/A_{\text{solid}}$\\
 &  (-) &  mm &  mm & mm$^2$ & (-)\\
\midrule
N-350  & 50    & 0.3  & 0.4  & 2.6  & 4\\
N-330  & 30    & 0.3  & 0.6  & 3.9  & 4\\
N-950  & 50    & 0.9  & 1.1  & 23.2 & 4\\
\bottomrule
\bottomrule
\end{tabular}
}
\end{table}

\subsection{Manufacturing of the AM nozzles}\label{sec:AM}

For the manufacturing of the nozzles, the powder bed fusion of metals using a laser beam (PBF-LB/M) process was employed. In this process, the part is manufactured incrementally by selectively fusing powder layers by a focused laser beam \cite{gibson2021}. All nozzle specimens were manufactured in a single pass on an EOS M400-1 PBF-LB/M machine (EOS GmbH, Germany) from the nickel-based superalloy Inconel~718 (Oerlikon Metco Europe GmbH, Germany). This machine was equipped with a 1~kW continuous-wave Yb fiber laser with a wavelength of 1064~nm. 
To minimize the influence of process-related effects, which were not investigated in this study, the specimens were arranged randomly on the build platform, which was kept at a constant temperature of 353~K during the manufacturing process. To maintain an inert atmosphere within the process chamber and prevent high-temperature oxidation of the molten metal, 99.999~\% pure Argon was used. While the geometry of the lattice structures was changed, all nozzle specimens were manufactured using the standard process parameters for Inconel~718. The laser power \textit{P} was set to 285~W, the scan speed \textit{v} to 960~mm/s, the hatch distance \textit{h} to 110~µm, and the layer thickness $d_{\text{L}}$ to 40~µm. A non-patterned line exposure with a rotation of 67° between two subsequent layers was selected as the scan pattern. The manufacturing process was followed by several post-processing steps, including the thorough cleaning of the specimens, the machining of the functional surfaces, and the drilling of holes for the integration of various sensors.

\begin{figure}
    \includegraphics{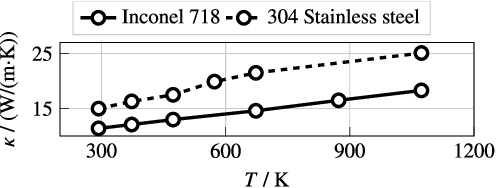}
    \caption{Thermal conductivity $\kappa$ of Inconel~718 (solid line) and 304 stainless steel (1.4301) (dashed line) with increasing temperature}
    \label{fig:thermal_conductivity}
\end{figure}

Since, in addition to the different lattice structures in the nozzle wall, the material properties also influence the combustion process, Fig.~\ref{fig:thermal_conductivity} illustrates the differences in the thermal conductivity $\kappa$ between the two tested nozzle materials as a function of the temperature~\textit{T}.

\subsection{Test rig}\label{sec:Combustor setup and test rig}

\begin{figure}
    \includegraphics{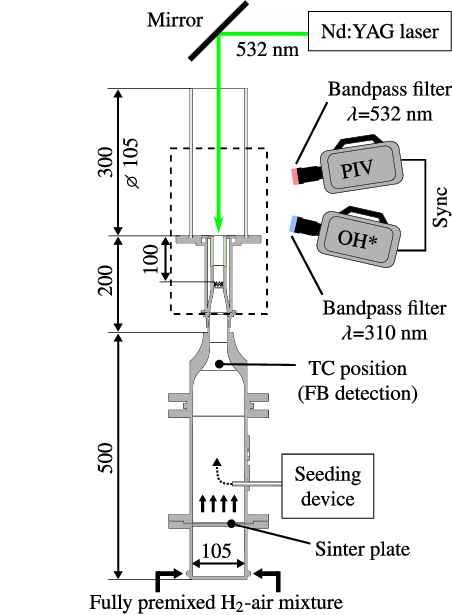}
    \caption{Test rig with the optical measurement setup. The flow direction is from bottom to top and all length scales are given in mm.}
    \label{fig:testrig}
\end{figure}

\begin{figure}
    \includegraphics{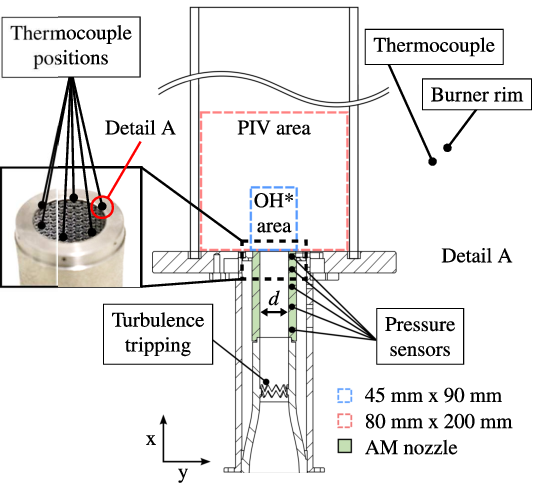}
    \caption{Detailed cross-section of the jet burner with the mixing duct and the changeable nozzle section of outlet diameter $d=20$~mm (highlighted in green). In Detail~A, the mounting situation of the thermocouple at the burner rim is depicted.}
    \label{fig:burner}
\end{figure}

The atmospheric jet flame test rig, shown in Fig.~\ref{fig:testrig}, comprised a premixed hydrogen-air mixture entering through six circumferential inlets at the burner base. Mass flow controllers regulated the fuel and air with an accuracy of $\pm$0.2~\%. The mixture passed through a metal sinter plate with a pore size of 25~$\mu$m to prevent flame FB into the supply lines, then entered the settling chamber where seeded air was added for the particle image velocimetry (PIV) measurements.
The AM mixing duct contained a thermocouple (TC) at its entrance to detect flame FB. A detected temperature rise by a flame moving unintentionally upstream into the mixing duct section triggered an automatic shutdown. 
A 1~mm thick zig-zag structure (see Fig.~\ref{fig:burner}) was positioned 100~mm upstream of the combustion chamber outlet to promote a rapid transition to a turbulent flow.

The flame anchored close to the mixing duct outlet inside the 300~mm long quartz glass combustion chamber with an open outlet. The inlet temperature of the unburnt mixture was 293~K.

The upper mixing duct section (indicated by the green area in Fig.~\ref{fig:burner}) was changeable and accommodated the different AM nozzles as described in Sec.~\ref{sec:nozzles}. All nozzles had an outlet diameter of $d=20$~mm. 
The burner flange had internal air cooling with a constant cooling mass flow of 8~kg/h and an inlet temperature of 293~K.

Five differential pressure sensors (1000~Pa range, 100~$\mu$s response time) were positioned along the mixing duct wall at a spacing of 0.5d -- 0.8d, starting 2.5~mm downstream of the combustion chamber. Daily zero-offset calibration was performed. Stable operating points were sampled at 15~kHz and FB points at 10~kHz both evaluated over 3~seconds. The power spectra were calculated using the Welch method with a Hanning window, a segment length of 2500 samples, and a window overlap of 50~\%. This resulted in a frequency resolution of 0.33~Hz. Savitzky-Golay filtering was utilized for noise reduction. The Nyquist frequency was set to half the sample rate.
Six type-K thermocouples with a diameter of 1~mm were flush-mounted circumferentially at the nozzle outlet, 1.2~mm upstream of the combustion chamber, shown in Fig.~\ref{fig:burner}. Thermocouple and mass flow data were acquired for 30~seconds at a sampling frequency of 10~Hz. \newline

\subsection{Operating conditions}\label{sec:operating conditions}

An overview of the operating conditions tested in this study is presented in Tab.~\ref{tab:operating conditions}.

\begin{table}[!ht]
\caption[Table]{Overview of the operating conditions}\label{tab:operating conditions}
\centering{%
\begin{tabular}{cccccc}
\toprule
\toprule
                          & $\dot{m}_{\text{air}}$ & $\bar{u}_{\text{x}}$ & $\phi$         \\
                          & kg/h                   & m/s                  & (-)            \\
\midrule
stable                    & 10; 13                 & 9; 11.5              & 0.39; 0.44; 0.48\\
FB test                   & 8; 10; 12              & 8.5; 9; 11           & 0.44 -- 0.6     \\
\bottomrule
\bottomrule
\end{tabular}
}
\end{table}

The air mass flows $\dot{m}_{\text{air}}$ under stable operating conditions were investigated for three equivalence ratios $\phi$ unless FB occurred beforehand, corresponding to calculated adiabatic flame temperatures $T_{\text{ad}}$ of 1400~K, 1500~K, and 1600~K. The thermal power ranged from 3.7~kW to 7~kW depending on the conducted air mass flow and equivalence ratio.
The flame properties were calculated using the \textsc{Cantera} chemical kinetics solver \cite{cantera2018goodwin} and the GRI~3.0 mechanism \cite{gri3}.
Reynolds numbers at the mixing duct outlet ranged from 9,000 -- 12,000 depending on the operating condition, calculated using the bulk velocity $\bar{u}_{\text{x}}$ of the unburnt hydrogen-air mixture and the nozzle outlet diameter $d$.

Flashback, defined as the upper stability limit of the burner where the flame unintentionally enters the mixing duct and moves upstream, was triggered using a controlled procedure. While the air mass flow remained constant, the flame temperature was increased in small increments $\Delta T_{\text{ad}}$ of 10 -- 15~\text{K} by raising the hydrogen mass flow. After each increment, the thermal equilibrium indicated by the temperature sensors was awaited before the next increase of the fuel mass flow.
When a FB occurred, the thermocouple inside the combustor detected the temperature rise from the upstream flame position, automatically shutting down the test rig. This feedback control loop ($\approx$2 seconds response time) prevented damage to the equipment. A cool down period of 10 minutes followed each shutdown to ensure identical initial conditions for each measurement.
The procedure achieved a high repeatability, with deviations of 0.64 -- 0.83~\% of the calculated adiabatic flame temperature which is within the range of the incremental temperature increase $\Delta T_{\text{ad}}$ used to trigger the FB. For technical reasons, the higher air mass flow of $\dot{m}_{\text{air}}=13$~kg/h had to be reduced to $\dot{m}_{\text{air}}=12$~kg/h for the FB tests with the procedure remaining the same. \newline

\subsection{Optical measurements}\label{sec:Optical measurements}

The PIV laser sheet entered the combustion chamber from the downstream end through a mirror angled at 90$^{\circ}$ above the outlet (Fig.~\ref{fig:testrig}), minimizing interfering reflections on the quartz glass walls. An air stream cooled the mirror and deflected hot exhaust gases to prevent overheating.
The image regions covered by the cameras are highlighted in Fig.~\ref{fig:burner}. 

The PIV system employed a double-pulse Nd:YAG laser ($\lambda=532$~nm, 10~kHz repetition rate) with a pulse separation of $\Delta t=100~\mu$s, which was kept constant across all measurements to ensure an adequate particle displacement between the snapshots. Zirconium dioxide particles ($d_{\text{ZiO}_2}=0.5~\mu$m) were seeded 600~mm upstream of the combustion chamber outlet by an seeding air mass flow of 2~kg/h, independent of the actual combustion air mass flow $\dot{m}_{\text{air}}$. The seeding mass flow was taken into account in the calculations.

Mie scattering was detected by a Photron~SA-1.1~CMOS high-speed camera with a 532~nm bandpass filter. The measurement area covered $80~\text{mm}\times 200~\text{mm}$ around the mixing duct outlet with a sheet thickness of 1.5~mm. 
The PIV snapshots were recorded with a resolution of 5.8~px/mm. For each measured case 23,000 snapshots were recorded that yield a record length of 2.3443~s. Post-processing employed standard PIV techniques with a $16~\text{px}$~$\times$~$16~\text{px}$ multi-grid interrogation window at 81~\% overlap resulting in $200\times166$ grid points for the calculated velocity field.

Flame imaging was performed using line-of-sight OH*-chemiluminescence measurements. The OH* radical provides a qualitative indication of the heat release zone, representing the flame front, as was shown by \cite{schiavone:2024}. 
An intensified Photron SA-Z~CMOS high-speed camera equipped with a bandpass filter with a wavelength of $\lambda=310$~nm recorded at 10~kHz with a spatial resolution of 7.8~px/mm. PIV and flame imaging was conducted synchronously. The imaging was triggered by the PIV camera for a temporal synchronization, which resulted in an identical acquisition time and number of snapshots. Mean flame images were acquired separately using a bandpass-filtered sCMOS camera with an exposure time of 0.8~s and a spatial resolution of 9.5~px/mm. \newline

%%% SPOD
\subsection{Spectral proper orthogonal decomposition}\label{sec:spod}

To identify coherent structures within the turbulent flow, spectral proper orthogonal decomposition (SPOD) was applied to the PIV snapshots. SPOD, unlike space-only POD, operates in the frequency domain to extract structures that are coherent both in space and in time, allowing for the analysis of flow dynamics. A more detailed explanation can be found in the work of~\cite{schmidt2020}.
Typically, SPOD is applied to the fluctuating component of a variable, so the mean is first subtracted from each snapshot before the analysis
\begin{equation}
    \mathbf{q}' = \mathbf{q}-\overline{\mathbf{q}},
\end{equation}
where $\mathbf{q}$ is the state vector at a given time instance and contains the information of the PIV snapshot. With the prime and overline symbol the fluctuating and the time-averaged component is denoted, respectively. In the following step, temporal blocks of the fluctuating quantity are Fourier-transformed. The Fourier amplitudes of all blocks are stacked in a matrix for the respective frequencies
\begin{equation}
    \widehat{\mathbf{Q}}_\omega = \left[\widehat{\mathbf{q}}_\omega^{(1) } \,\,\widehat{\mathbf{q}}_\omega^{(2) } \,\,\widehat{\mathbf{q}}_\omega^{(3) } \,\, ... \,\, \widehat{\mathbf{q}}_\omega^{(N_{\text{B}}) } \, \right],
\end{equation}
where $\widehat{\mathbf{q}}_\omega^{(j) }$ denotes the complex Fourier amplitude at the temporal frequency $\omega$ obtained from block $j$. The number of blocks is denoted $N_{\text{B}}$ and depends on the length of the time series, the length of the individual blocks and their selected overlapping. Based on $\widehat{\mathbf{Q}}_\omega$, the cross spectral density matrix is determined, whose eigenvectors are the SPOD modes  
\begin{align}
    \frac{1}{N_{\text{B}} -1} \widehat{\mathbf{Q}}_\omega \widehat{\mathbf{Q}}_\omega^H \widehat{\Psi}_\omega  = \widehat{\Psi}_\omega \widehat{\Lambda}_\omega.
\end{align}
The superscript $H$ denotes the complex conjugate transpose. The columns of $\widehat{\Psi}_\omega$, denoted $\widehat{\psi}_\omega^{(j)}$ are the SPOD modes and the elements of the diagonal matrix $\widehat{\Lambda}_\omega$, denoted $\lambda_\omega^{(j)}$ are the corresponding SPOD eigenvalues.
The SPOD modes are orthonormal, and for each mode, the energy is given directly by its eigenvalue. High‑energy modes are particularly relevant for the flow dynamics and the modes are therefore ordered by decreasing eigenvalue such that the first mode, with eigenvalue $\lambda_\omega^{(1)}$, is the most energetic. If this leading mode contains substantially more energy than the subsequent modes, the spectrum is said to exhibit low‑rank behavior, meaning that the dynamics at that frequency are dominated by the first mode. 

In the following, the axial and transverse velocity components from PIV were analyzed with SPOD. For the velocity, the energy is given in units of m$^2$/s$^2$ and can be interpreted as a measure for the turbulent kinetic energy, neglecting density variations. For brevity, only the real part of the SPOD modes is shown in the following. To ensure that the amplitude is representative, the real part is displayed at the phase angle at which it was largest.

The length of the PIV time series for the SPOD was reduced to 12,800 snapshots to provide a consistent image number for both stable and unstable operating conditions. To focus the calculation on the flow field region impacted by the flame, the snapshots were cropped in the downstream direction at $x/d=3.9$. For the decomposition, the number of frequencies was set to $n_{\text{DFT}}=256$ with an overlap of 50~\%, resulting in 49 blocks for the POD. \newline

%%%%%%%%%%%%%%%%%%%%%%%%%%%%%%%%%%%%%%%%%%%%%%%%%%%%%%%%%%%

\section{Experimental results}\label{sec:Results}
In the following section, the results of the measurements are presented and analyzed. First, the mean flow was analyzed for effects caused by changing the wall structure. Subsequently, the FB behavior was investigated, and finally, the flow dynamics were examined.

\subsection{Mean flow}\label{sec:mean_flow}
\begin{figure}
\centering
    \includegraphics{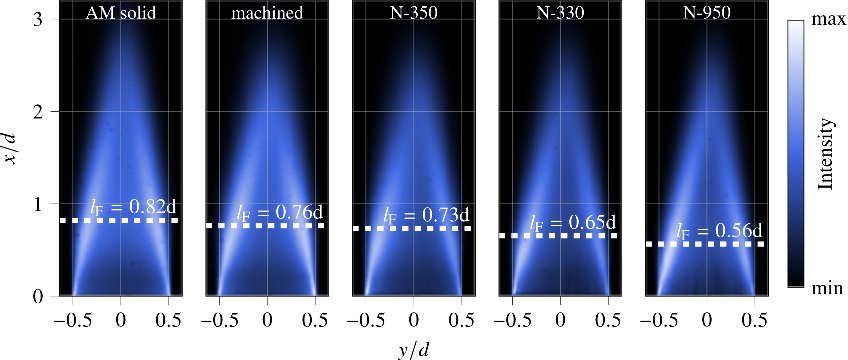}
    \caption{OH*-chemiluminescence flame images at an adiabatic flame temperature of $T_{\text{ad}}=1600$~K and an inlet velocity of $\bar{u}_{\text{x}}=11.5$~m/s. The flame length $l_{\text{F}}$ was determined by the maximum of the mean intensity along the $y$-axis.}
    \label{fig:flameimages}
\end{figure}

Figure~\ref{fig:flameimages} shows the averaged flame images of the line-of-sight integrated OH* signals for the five nozzle variants. The operating point was at an adiabatic flame temperature of $T_{\text{ad}}=1600$~K and a bulk flow velocity of $\bar{u}_{\text{x}}=11.5$~m/s.

For all cases, the flames exhibited a typical conical jet shape, with increased OH* intensity along the cone edges due to the line-of-sight integration. The axial position of the flame length $l_{\text{F}}$ was defined by the maximum of the OH* mean intensity integrated over the $y$-direction. The strongest heat release occurred near the calculated flame length and is accompanied by a thickening of the edge region with high OH* intensity. 
The length of the flame tends to decrease as the lattice cell size increases (left to right in Fig.~\ref{fig:flameimages}). A notable decrease in the flame length was observed for the nozzle N-950, which has the coarsest lattice structure. This may result from an increased turbulence by the lattice structures of the walls.
Nozzles with porous walls show a slight asymmetry in the heat release in Fig.~\ref{fig:flameimages}. Evaluated over all measured cases this asymmetry depends on the operating condition and, to some extent, also varied for the nozzles with solid walls.

\begin{figure}
    \includegraphics{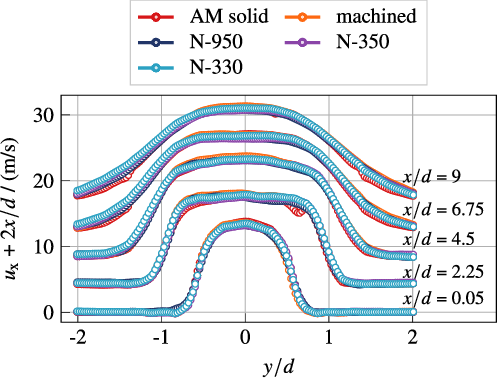}
    \caption{Axial velocity component $\bar{u}_{\text{x}}$ shown at the axial positions $x/d$ in the downstream direction over the combustion chamber width $y/d$. The operating condition was at an adiabatic flame temperature of $T_{\text{ad}}=1600$~K and an inlet velocity of $\bar{u}_{\text{x}}=11.5$~m/s.}
    \label{fig:axial_velocity}
\end{figure}

To address the observed changes in the shape of the flames, Fig.~\ref{fig:axial_velocity} shows the axial mean velocity $\bar{u}_{\text{x}}$ at different downstream positions in the combustion chamber. 
The operating conditions were identical to those of the OH* images in Fig.~\ref{fig:flameimages}. The velocity profiles exhibited the typical decay and radial spreading expected for a jet configuration. Apart from small deviations, all measured profiles agreed well and showed comparable velocities.

At $x/d=2.25$, an increase in the velocity was observed, which can be attributed to the flame position and the associated decrease of the density over the flame front. Toward the combustion chamber wall, the velocity decreased almost to zero. With an area ratio of 25:1 between the burner plate area and the mixing duct outlet area, the burner generated only a very weak recirculation zone in the lower part of the combustion chamber (not shown). The recirculation strengthens once the jet has extended sufficiently to form vortices between the jet and the combustion chamber wall at approximately $x/d=5$. Nevertheless, the recirculation extended in a narrow band along the shear layer down to the flame root.

\begin{figure}
    \includegraphics{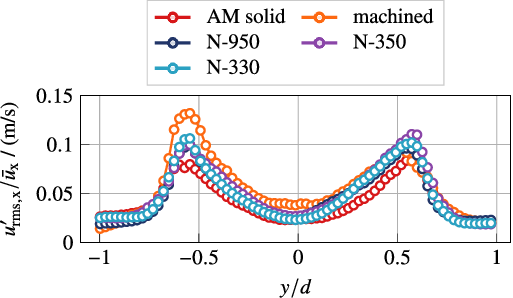}
    \caption{Turbulence intensity of the axial velocity component downstream of the nozzle outlet at $y/d=0.05$ shown at the same operating condition as in Fig.~\ref{fig:axial_velocity}.}
    \label{fig:axial_velocity_fluctuation}
\end{figure}

The turbulence intensity at the nozzle outlet is shown in Fig.~\ref{fig:axial_velocity_fluctuation}, with color coding identical to the previous figure for comparison. The fluctuation velocity $u^{\prime}_{\text{rms}}$ was normalized by the spatial mean of the axial velocity $\bar{u}_{\text{x}}$, with the evaluation restricted to the nozzle outlet region.

The highest turbulence intensities were measured in the shear layer regions near the nozzle edges ($y/d=\pm0.5$), decreasing toward the nozzle center. Small differences appeared among the nozzle variants. The machined nozzle showed a high, non-symmetric peak at $y/d=-0.5$, with $u^\prime_{\text{rms}, \text{x}}/\bar{u}_{\text{x}}=0.113$, which also represents the largest asymmetry observed. The other nozzles peaked at values ranging from 0.08 -- 0.10, consistent with expected values for these experimental conditions \cite{fellouah:2009}. 
It is concluded that, in contrast to the flame length, no trend related to the lattice structures is observed for the flow field.\newline

\subsection{Flashback behavior}\label{sec:Flashback behavior}
After investigating the mean flow under stable flame conditions in the previous section, the focus now shifts to the FB limit of the burner setups. 

\begin{figure}
    \includegraphics{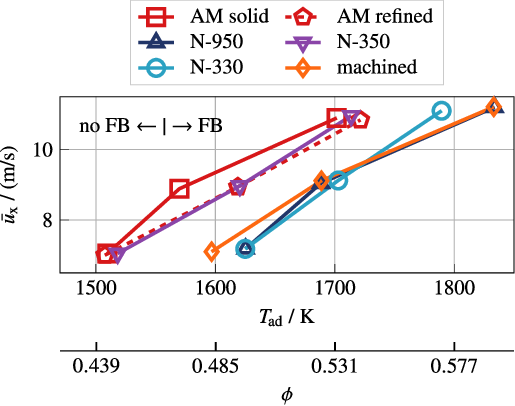}
    \caption{Flashback limits with the inlet velocity of the unburnt gas mixture $\bar{u}_{\text{x}}$ as a function of the calculated adiabatic flame temperature, or equivalence ratio. A flashback is indicated by passing the linear interpolated solid lines from left to right.}
    \label{fig:flashback_limits}
\end{figure}

Figure~\ref{fig:flashback_limits} shows the FB limits, determined using the procedure described in Sec.~\ref{sec:operating conditions}. The thresholds are represented by the axial flow velocity as a function of the calculated adiabatic flame temperature, or equivalence ratio, calculated from the last measured mass flows before a FB occurred. Depicted are the five nozzle variations as described in Sec.~\ref{sec:nozzles} as well as a post-processed version of the solid AM nozzle referred to as \textit{AM~refined}, indicated by the red dashed line with the pentagon symbols in Fig.~\ref{fig:flashback_limits}.
For a better visualization of the trend in the FB data, a linear interpolation between the measured data points was applied. To the left of each interpolated line, stable flame conditions can be assumed. When the flame temperature is increased or the inlet velocity is reduced so that the line is crossed from left to right, the FB limit is reached. All observed FB events were boundary layer FB, confirmed by high-speed imaging.

The highest FB propensity was found for the solid AM nozzle, followed by the nozzle N-350, which had the smallest cell size (finest porous structure). The upper FB limit was shared by the machined nozzle and the nozzle N-950, the latter having the largest cell size of the investigated nozzles. The nozzle N-330 also showed a high FB resistance, but performed worse at a bulk velocity of $\bar{u}_{\text{x}}=11$~m/s.
Therefore, Fig.~\ref{fig:flashback_limits} shows two distinct trends in the FB resistance:
\begin{itemize}
    \item from low to high thermal conductivity for the solid nozzles with identical geometry
    \item from small to large cell sizes for the nozzles comprising lattice structures 
\end{itemize}
First, the large gap between the FB limits of the two solid nozzles was investigated. Because the inner surface of the solid AM nozzle was not post-processed, it exhibits the typical surface roughness of this manufacturing method. Typical literature values for the average surface roughness of the AM nozzle material Inconel~718 are $R_a\approx9~\mu$m compared to $R_a\approx1~\mu$m for stainless steel \cite{psomoglou:2024}.
The surface roughness can have a very different influence on the FB tendency depending on the material used. 
Ding et al.~\cite{ding2017} conducted a numerical study that observed a decreasing critical velocity gradient with greater wall roughness. This is particularly significant when the material of the wall has a high thermal conductivity, as it results in an increased flame quenching distance due to a greater heat loss. However, flow field modifications in the near-wall region can counteract the effect. Vivoli et al.~\cite{vivoli:2025} reported experimental data showing a strong decrease of the FB risk with the increase of the surface roughness utilizing stainless steel mixing tubes.

To exclude the surface roughness as the cause of the observed differences between the two solid nozzles, the solid AM nozzle was post-processed on a lathe to obtain a surface roughness comparable to that of the machined stainless steel nozzle. 
The reduced surface roughness of the refined AM nozzle slightly increased the FB resistance compared to the solid AM nozzle without treatment of the inner mixing tube wall (see Fig.~\ref{fig:flashback_limits}) but this effect was too weak to explain the previously observed gap in the FB risk. It is noted, given the thermal conductivity of Inconel~718 under the tested conditions, that the surface roughness resulting from the manufacturing process is negligible in terms of decreasing the FB propensity.

Since surface roughness effects had been ruled out, the primary difference between the two solid nozzles was the material. Inconel~718 has a thermal conductivity of $\kappa=11.4$~W/(m$\cdot$K), which is 25~\% lower than the thermal conductivity of 304 stainless steel with $\kappa=15$~W/(m$\cdot$K), both at $T=293$~K (see Fig.~\ref{fig:thermal_conductivity}).
Duan et al.~\cite{Duan:2014} have reported that the thermal conductivity strongly influences the FB limits. However, in their cases (brass, quartz glass, and steel), the differences in the thermal conductivity were spanning over one order of magnitude, providing only a qualitative trend relevant to the present study.

\begin{figure}
    \includegraphics{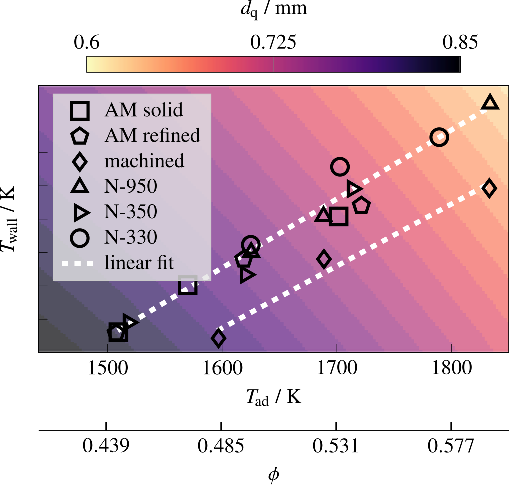}
    \caption{Measured wall temperatures (black marker) at the mixing duct outlet as a function of the calculated adiabatic flame temperature, or equivalence ratio. All measurement points are close to the FB condition. The background color depicts the associated quenching distance $d_{\text{q}}$, calculated based on \cite{kiymaz:2025}.}
    \label{fig:quenchingDistance}
\end{figure}

To assess the effect of the thermal conductivity on the FB limit, the wall temperature $T_{\text{wall}}$ near the mixing duct rim was evaluated, shown in Fig.~\ref{fig:quenchingDistance}. At stable operating conditions, no clear trend in the wall temperature was observed, which is not shown for brevity. 
Near FB, the flame anchors closer to the burner rim and exerts the strongest influence on the burner rim temperature. Therefore, the temperature measurements immediately before FB showed distinct separation according to the material of the nozzle.
As indicated by the white dashed lines in Fig.~\ref{fig:quenchingDistance}, the data points of all Inconel~718 nozzles collapse on one line (upper white dashed), while the flashback points of the stainless steel nozzle form another line (lower white dashed). 
Therefore, it is concluded that the difference in the thermal conductivity caused the observed gap in the FB limits between the solid AM nozzle and the machined nozzle.

Next, the observed trend of the FB limits related to the pore size of the lattice structures is evaluated.
In Fig.~\ref{fig:quenchingDistance}, the quenching distance $d_{\text{q}}$ of the nozzles is shown by the background color coding. The quenching distance was calculated using an empirical formula for hydrogen-air mixtures derived by K\i ymaz et al.~\cite{kiymaz:2025}, which incorporates the wall temperature $T_{\text{wall}}$ and the equivalence ratio. The quenching distance exhibits a linear trend as a function of the wall temperature, observed in other studies as well \cite{kosaka2018,zirwes2021}. 
It is noted in Fig.~\ref{fig:quenchingDistance} that the calculated quenching distances of the AM nozzles with lattice structure were lower compared to the machined nozzle at the same flame temperature, indicating a higher FB resistance despite the lower thermal conductivity.

Under the present conditions, three potential mechanisms driving the observed FB behavior of the nozzles with lattice structure were identified and will be discussed in the remainder of this study:
\begin{enumerate}%[\quad (1)]
    \item[\quad (1)] alteration of the mean fields with decreased critical velocity gradient
    \item[\quad (2)] passive wall cooling by fresh gases flowing through the lattice structure
    \item[\quad (3)] modification of flow dynamics damping flame instability
\end{enumerate}

\noindent With regard to the first potential mechanism, it should be noted that the mean flow field data in Sec.~\ref{sec:mean_flow} showed no significant differences between the nozzle variants. For the flame topology, a trend related to pore size was observed where the flame length decreased as pore size increased. Typically, a decreased flame length at identical operating conditions implies an increased turbulent burning velocity, which would theoretically increase FB propensity~\cite{hoferichter2017}. However, the opposite was observed here as the FB limit actually increased with shorter flame lengths (and larger pore sizes). This suggests that alteration of the mean fields was not the primary driver of the improved flame stability and rather destabilized the flame.

Instead, the trend in FB limits shown in Fig.~\ref{fig:flashback_limits} directly correlates with the pore size. The lattice structures can significantly enhance the convective heat transfer capability of the wall by the strongly increased surface area. Therefore, a passive wall cooling by the unburnt fresh gases might mitigate FB.

\begin{figure}
    \includegraphics{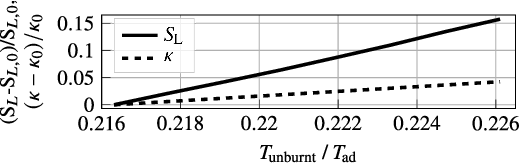}
    \caption{Effect of an inlet temperature variation ($\Delta T_{\text{unburnt}}=20$~K) of the unburnt gas mixture on the laminar flame speed $S_{\text{L}}$ and the thermal conductivity of unburnt gas $\kappa$. Both quantities are normalized by their respective initial conditions $S_{\text{L},0}$ and $\kappa_{0}$ at $T_{\text{unburnt}}=360$~K.}
    \label{fig:flamespeedIncrease}
\end{figure}

An increased heat transfer of the wall influences the temperature of an unburnt gas mixture $T_{\text{unburnt}}$ near the mixing duct wall \cite{kosaka2018} and thus the state of the flame reactants in proximity to the lattice structures. The results from one-dimensional free flame simulations are presented in Fig.~\ref{fig:flamespeedIncrease}. Shown is the unstretched laminar flame speed $S_{\text{L}}$ and the thermal conductivity $\kappa$ of the fresh gases as a function of the temperature ratio $T_{\text{unburnt}}$ / $T_{\text{ad}}$, respectively. Both values are sensitive to changes in the unburnt gas mixture temperature with a decreasing wall temperature resulting in an increasing quenching distance.

Studies by \cite{ding2017,vivoli:2025} demonstrated that an increased heat transfer at the wall correlates with FB reduction.
However, the penetration depth and turbulence at the cell level are difficult to quantify for the lattice structures. A rigorous investigation would require using numerical simulation for resolving or modeling the porous wall structure, conjugate heat transfer, reaction modeling, and turbulent flow at $\text{Re}=10,000$, which is outside the scope of this study.
However, investigations on the conjugate heat transfer of lattice structures at a comparable Reynolds number and porosity to nozzle N-950 in a rectangular duct, were conducted by Kuwata et al.~\cite{kuwata:2020} and Suga et al.~\cite{suga:2020}. In their numerical and experimental studies, they observed penetration of the fluid into the porous media, accompanied by an increased wall turbulence and an enhanced heat transfer. The rise in turbulence was attributed to Kelvin-Helmholtz (KH) instability developing over the porous wall.
Thereby, the nozzle end plate in the present setup is crucial for preventing a reduction in the exit velocity and for supporting flame anchoring, as confirmed by tests with nozzle variants without an end plate, which is not shown for brevity.

As a result, the second hypothesis, i.e. wall cooling by the fresh gases flowing through the lattice structures appears to be a plausible cause for the improved flashback behavior. 

Besides the effect of wall cooling, the motion of the fluid through the lattice structures might effectively reduce the Soret effect, which is considered to be an important factor for FB in hydrogen combustion \cite{porath:2025,fruzza:2024}.

The following sections address the third potential mechanism of whether lattice structures modify the flow dynamics to impact the FB behavior or the flow field. Starting point are the flow dynamics in the mixing duct.   \newline

\subsection{Mixing duct flow}\label{sec:pressure measurements}
Numerical studies by Wang et al.~\cite{wang:2022} have demonstrated that flow instabilities in the mixing duct region of jet flame burners influence the resulting flow field and flame stability. Garnaud et al.~\cite{garnaud:2013} reported the origin of flow structures in the wall boundary layer of a non-reacting jet. 

\begin{figure}
    \includegraphics{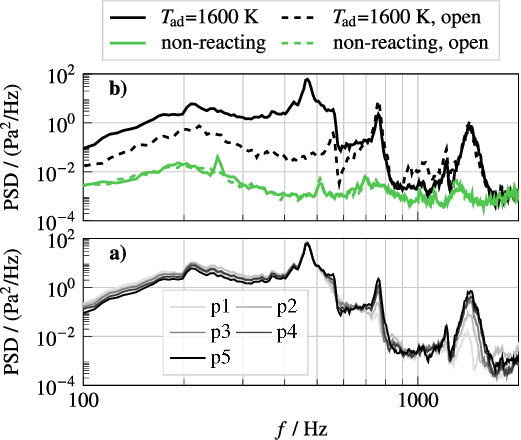}
    \caption{PSD of the machined nozzle at $\bar{u}_{\text{bulk}}=9$~m/s. The bottom plot shows the signals of the five axial measurement positions (p1 to p5). Line brightness decreases with increasing distance to the mixing duct outlet. The top plot shows the PSD of pressure sensor p5 for the non-reacting cases (green lines) and the cases with flame at $T_{\text{ad}}=1600$~K (black lines) both with (solid lines) and without (dashed lines) confinement.} 
    \label{fig:powerSpectra_pressure_overview}
\end{figure}

Since optical access is not possible with the lattice structures, pressure measurements were used to assess effects related to the lattice structures and the onset of FB.
Figure~\ref{fig:powerSpectra_pressure_overview}~a) shows the power spectral density (PSD) calculated for the five pressure sensors p1 to p5 along the mixing duct.

The power spectra of the five sensors show the amplitude decreasing from upstream (p5) to downstream (p1), indicating a pressure node at the outlet of the mixing duct that might be related to a thermo-acoustic (TA) mode. No significant differences appeared across the different axial positions. Therefore, the subsequent analysis was focused on the sensor p5.

Reacting conditions at $T_{\text{ad}}=1600$~K, shown in Fig.~\ref{fig:powerSpectra_pressure_overview}~b) by the black lines, displayed significant higher signal strength and altered spectral shape compared to non-reacting conditions indicated by the green lines.
Higher frequencies (300 -- 2000~Hz) were more promoted with flame and three significant peaks emerged at 468~Hz, 762~Hz, and 1440~Hz. A comparison with data at reacting conditions but without a quartz glass confinement (black dashed line) showed that reflections from the combustion chamber wall increased the spectral baseline for frequencies below approximately 600~Hz, making the peak at 468~Hz a possible TA mode. Consequently, the two peaks at 762~Hz and 1440~Hz were identified as flame-associated modes. 

To assess the influence of the lattice structures, a stable operating point ($T_{\text{ad}}=1500$~K, dashed lines) and an unstable point near FB ($T_{\text{ad}}=1690$~K, solid lines) are compared in Fig.~\ref{fig:powerSpectra_pressure_FB} for the nozzle N-950 and the machined variant.
Both followed identical trends for the two flame conditions. With an increasing flame temperature, low-frequency pressure amplitudes decreased while high-frequency amplitudes increased. The frequency shift likely resulted from density changes at elevated temperatures and the related change in the speed of sound. This behavior was observed for all nozzle variants, but to varying degrees. A more comprehensive analysis can be found in App.~\ref{sec:appendix}.

It can be concluded, that the pressure fluctuations in the mixing duct showed a distinct spectra with a peak at 468~Hz related to a TA mode and two peaks at 762~Hz and 1440~Hz related to the combustion process. Only the amplitudes of the latter ones were sensitive to flame temperature variations. An examination of the flow dynamics inside the combustion chamber and the influence of the TA mode on the combustion process are presented in the next section. \newline

\subsection{Flow dynamics in the combustion chamber}\label{sec:Flow dynamics}
After examining the non-reacting flow in the mixing duct, the reacting flow in the combustion chamber was analyzed and assessed whether these flow dynamics influence FB beyond the assumed effect of wall cooling.

\begin{figure}
    \includegraphics{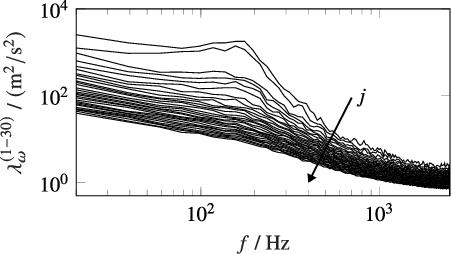}
    \caption{SPOD modes $j=1$ -- $30$ calculated from the PIV data. Shown is the machined case for $\bar{u}_{\text{x}}=~11.5$~m/s at a flame temperature of $T_{\text{ad}}=1600$~K.}
    \label{fig:SPOD_full_e0071}
\end{figure}

Figure~\ref{fig:SPOD_full_e0071} shows the first 30 SPOD eigenvalues for the machined nozzle at $T_{\text{ad}}=1600$~K and inlet velocity $\bar{u}_{\text{x}}=11.5$~m/s. 
A clear energy separation between the first two eigenvalues and the other modes was observed at frequencies of 100 -- 200~Hz, corresponding to Strouhal numbers $\text{St}=(f*d)/\bar{u}_{\text{x}}$ of $\text{St}=0.2$ -- 0.4, based on the given nozzle diameter. The two leading modes exhibited comparable energy levels dominating the low-frequency domain, suggesting a classification as rank two. After peaks at 175~Hz, a monotonic decrease in the eigenvalue magnitude was observed for all modes ($j=1$ -- 30), indicating that low-frequency fluctuations dominated the flow field.

\begin{figure}
    \includegraphics{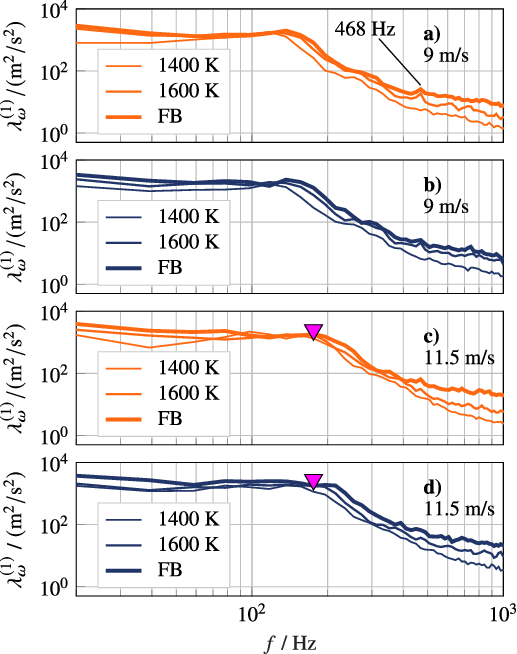}
    \caption{Leading SPOD modes for the machined nozzle (orange lines) and the nozzle N-950 (blue lines). FB shows the measurement point directly before flashback at the reduced bulk velocity of 11~m/s (thickest lines). Markers indicate the frequencies of the mode plots in Fig.~\ref{fig:SPOD_e0092_mode1_195Hz}.}
    \label{fig:SPOD_subplot_velocities}
\end{figure}

Since low-rank behavior of the two leading modes was observed across all nozzle variants, the analysis focused on the machined nozzle and the nozzle N-950. Figure~\ref{fig:SPOD_subplot_velocities} shows the leading eigenvalues $\lambda_\omega^{(1)}$ at adiabatic flame temperatures of 1400~K, 1600~K and directly before FB at $T_{\text{ad}}=1840$~K. 
Panels a) and c) present the machined nozzle at inlet velocities of approximately 9~m/s and 11.5~m/s for stable conditions, respectively, while panels b) and d) show the same velocities for the nozzle N-950. For the unstable FB condition, the velocity was reduced to 11~m/s as described in Sec.~\ref{sec:operating conditions}.
Because of the low mode energy above $f=1000$~Hz observed in Fig.~\ref{fig:SPOD_full_e0071}, the spectrum is truncated at this frequency to obtain a more detailed resolution.

Low-frequency dynamics dominated all SPOD spectra. With an increasing velocity, the low-frequency peak shifted to higher frequencies and became more broadband, while the energy content remained on a comparable level. In the high-frequency region, the increased velocity and flame temperature produced a less steep eigenvalue decay. This trend matches observations in the pressure spectra measured upstream in the mixing duct, see Fig.~\ref{fig:powerSpectra_pressure_FB}.

At $\bar{u}_{\text{x}}=9$~m/s, a distinct peak appeared at 468~Hz, corresponding to a TA mode identified in the pressure measurements in Sec.~\ref{sec:pressure measurements}. This mode was less pronounced in the lattice structure nozzle, which exhibited greater energy separation in the high-frequency region. This difference may result from increased small-scale turbulence generated by the nozzle wall structure which would match the observed reduction of the flame length in Fig.~\ref{fig:flameimages}. 
However, the observed TA mode was very weak and showed no significant effect in the heat release spectra of the flame, an influence on the FB limits at the present operating conditions was assumed to be negligible.
Overall, SPOD eigenvalues displayed broader frequency distributions with less pronounced peaks than the corresponding pressure spectra in the mixing duct.

\begin{figure}
    \includegraphics{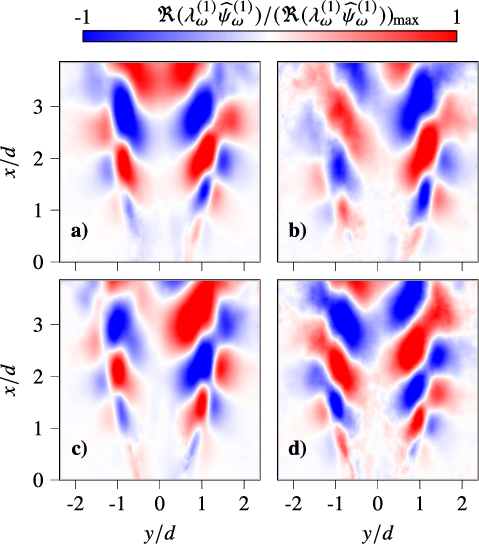}
    \caption{Real part of the leading SPOD mode at $f=171$~Hz indicated in Fig.~\ref{fig:SPOD_subplot_velocities}. Panels a) and c) show the stable operating points and panels b) and d) the moment directly before flashback.}
    \label{fig:SPOD_e0092_mode1_195Hz}
\end{figure}

Figure~\ref{fig:SPOD_e0092_mode1_195Hz} illustrates the mode structure, showing the real part of the $j=1$ eigenmode at $f=171$~Hz (marked by triangles in panels c) and d) of Fig.~\ref{fig:SPOD_subplot_velocities}). The left panels, a) and c), present data at $T_{\text{ad}}=1600$~K, while the right panels, b) and d), show conditions directly before the FB at $T_{\text{ad}}=1840$~K.

With a Strouhal number of $\text{St}=0.3$, the mode reconstruction in panel a) reveals the axisymmetric component of a Kelvin-Helmholtz instability, while the non-symmetric component is located in the $j=2$ sub-leading mode (not shown). For the nozzle N-950 under identical conditions, the non-symmetric part projected more strongly onto the leading mode. The KH instability originated at the mixing duct outlet, with multiple phase reversals occurring along the shear layer in the downstream direction.

An increasing heat release toward higher flame temperatures enlarged the jet opening angle, particularly at the flame root near the mixing duct outlet (panels b) and d) near FB). This correlates with a stronger and more extensive recirculation, covering a larger fraction of the combustion chamber and extending toward the burner plate. Enhanced recirculation established an increased heat flux toward the burner plate, additionally explaining the pronounced increase in the burner rim temperature approaching FB conditions (Fig.~\ref{fig:quenchingDistance}). 

The results indicate that the flow dynamics in the combustion chamber were dominated by large-scale coherent structures in the shear layer, specifically KH instabilities. This corresponds to the observation made for the low-frequency range in the mixing duct flow with the increasing shear layer thickness affecting the FB behavior. As the flame temperature increased, the KH mode shifted to slightly higher frequencies and the energy contained in high-frequency modes grew. With the porous wall structure, turbulence was further enhanced, while a TA mode observed at $f=468$~Hz was damped.
However, a specific influence of the lattice structures on the flow dynamics that affects the FB propensity was not observed. \newline

%%%%%%%%%%%%%%%%%%%%%%%%%%%%%%%%%%%%%%%%%%%%%%%%%%%%%%%%%%%

\section{Conclusion}

This study demonstrates that incorporating BCC lattice structures into nozzles mitigates flame flashback in a hydrogen jet burner and further investigates the mechanisms underlying this effect. A comprehensive experimental characterization was conducted, combining flow measurements, flame imaging, and spectral analysis. Based on the observations, it is concluded that the primary mitigation mechanism is passive wall cooling by fresh gases flowing through the lattice structures with the cell size being related to this effect. Other potential mechanisms for the observed FB behavior were thoroughly discussed and ruled out based on experimental findings and literature data.

Analysis of the temperature data confirmed that wall temperature was the controlling parameter governing the FB limits across the different investigated nozzle configurations. The AM nozzle N-950 achieved a FB resistance significantly higher than the solid-wall AM nozzle and comparable to a reference nozzle with 25\% higher thermal conductivity. This demonstrated that lattice structure optimization enables a broader operating range. Following the findings of \cite{ding2017, vivoli:2025}, that an increased heat transfer of the wall correlates with a decreased FB risk, it is reasonable to assume that the presumed mechanism of passive wall cooling is transferable to other nozzle materials.

Flow field measurements showed only minor modifications resulting from incorporating lattice structures in the nozzle wall, preserving general jet characteristics while in the flame images a reduction of the flame length with increasing pore size was observed. SPOD revealed dynamic differences, with the Kelvin-Helmholtz instability dominating low-frequency dynamics, increased asymmetry in the lattice-structured nozzle, and an enhanced turbulence generation.

The findings confirmed that the integration of lattice structures through AM provides a viable strategy for hydrogen FB mitigation by manipulating the coupled interaction between the flame and the  thermal conditions of the wall. Future work should address the transferability to gas turbine relevant conditions in a multi-nozzle system, where elevated temperatures and flame-to-flame interactions may alter single-nozzle observations.
Furthermore, the systematic optimization of the lattice structure in relation to operating conditions, combined with thermal management strategies, could further enhance the FB resistance.

%%%%% Acknowledgments %%%%%%%%%%%%%%%%%%%%%%%%%%%

\section*{Acknowledgments}
This work has been conducted as part of the joint research project ENERGIZE (Adjoint-based and additive manufacturing-enabled optimization of hydrogen combustion systems) within the scope of the SPP2419 HyCAM and is supported by the German Research Foundation (DFG) under grant numbers 523881008, 530442286. 
Furthermore, the authors would like to thank Stefan Buchwald for his support at the measurements.

%%%%% Appendix %%%%%%%%%%%%%%%%%%%%%%%%%%%

\newpage
\appendix
\section{Pressure fluctuations in the mixing duct flow}\label{sec:appendix}
\begin{figure}
    \includegraphics{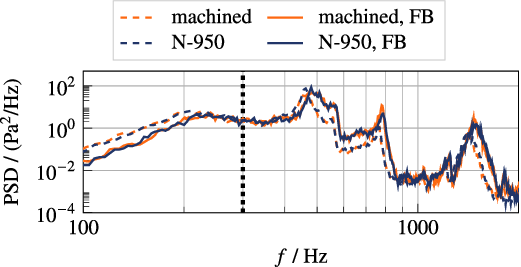}
    \caption{Power spectra of the machined nozzle and the nozzle N-950 at a bulk velocity of $\bar{u}_{\text{x}}=9$~m/s. Both cases are depicted at $T_{\text{ad}}=1500$~K (dashed lines) and $T_{\text{ad}}=1690$~K (solid lines). The latter measurement point is measured over the last three seconds before a FB occurred. The dotted black line indicates $f=300$~Hz.}
    \label{fig:powerSpectra_pressure_FB}
\end{figure}

To isolate the effects of the bulk velocity and the flame temperature on the pressure fluctuations $p^\prime$ in the mixing duct, the pressure fluctuations were averaged and normalized by the total energy $p^\prime_{\text{max}}$ in both the low-frequency (0 -- 300~Hz) and the high-frequency (300 -- 2000~Hz) ranges, shown in Fig.~A\ref{fig:powerPressure_subplot}. The frequency separation at $f=300$~Hz is based on the observations made in Fig.~\ref{fig:powerSpectra_pressure_FB}. The flame temperature is indicated by the color coding. 

Results for the stable operating points ($T_{\text{ad}}=1500$~K) appear in the panels a) and c), while the unstable points ($T_{\text{ad}}=1690$~K) are shown in the panels b) and d). 
With an increase of the flame temperature the pressure amplitude decreases in the low-frequency range and increases in the high-frequency range, indicating the frequency shift observed in Fig.~\ref{fig:powerSpectra_pressure_FB}. The intense of pressure fluctuations might be affected by the wall turbulence and the flame temperature. This effect is least prominent for the solid AM nozzle with the smallest difference of pressure amplitude between both frequency ranges.
%%%  REFERENCES  %%%%%%%%%%%%%%%%%%%%%%%%%%%%%%%%
\newpage

%%%  LIST OF FIGURE CAPTIONS  %%%%%%%%%%%%%%%%%%%%%%%%%%%%%%%%
\clearpage

\setcounter{figure}{0}
\renewcommand{\figurename}{Figure A}
\begin{figure}
    \includegraphics{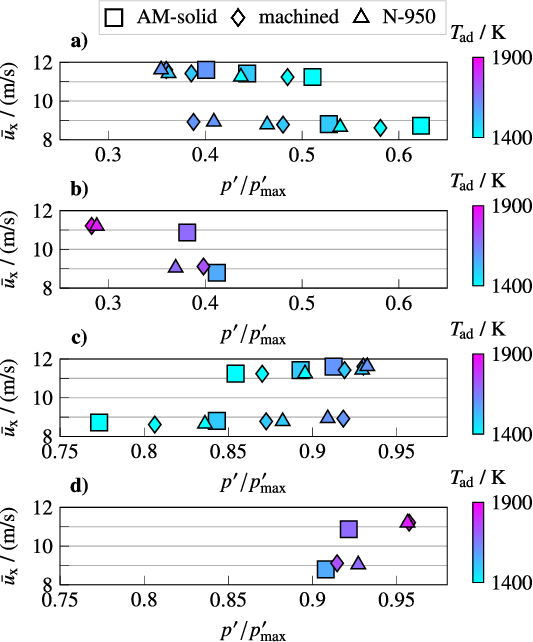}
    \caption{Normalized pressure amplitudes for the two spectra regions below and above $f=300$~Hz. Panels a) and b) show the low-frequency range (0 -- 300~Hz)) of the pressure fluctuations and panels c) and d) show the high-frequency range (300 -- 2000~Hz) of the pressure fluctuations. Panels a) and c) are at stable operating conditions and panels b) and d) close to FB.}
    \label{fig:powerPressure_subplot}
\end{figure}

%%%%%%%%%%%%%%%%%%%%%%%%%%%%%%%%%%%%%%%%%%%%%%%%%%%%%%%%%%%%%%%%%%%%%%%%%%%%%%%%%%%%%%%
\clearpage
\printbibliography

@Article{Ciani2019,
journal={Journal of the Global Power and Propulsion Society},
volume={3},
year={2019},
title={Superior fuel and operational flexibility of sequential combustion in Ansaldo Energia gas turbines},
author={Ciani, Andrea
and Bothien, Mirko R.
and Bunkute, Birute
and Wood, John P.
and Früchtel, Gerhard},
doi={10.33737/jgpps/110717},
}

@article{magnusson:2020,
    author = {Magnusson, Rikard and Andersson, Mats},
    title = "{Operation of SGT-600 (24 MW) DLE Gas Turbine With Over 60 \% H2 in Natural Gas}",
    volume = {Volume 9: Oil and Gas Applications; Organic Rankine Cycle Power Systems; Steam Turbine},
    series = {Turbo Expo: Power for Land, Sea, and Air},
    year = {2020},
    month = {09},
    doi = {10.1115/GT2020-16332},
}

@article{cosic:2022,
    author = {Ćosić, Bernhard and Wassmer, Dominik and Kluß, David and Jaeschke, Alexander and Reichel, Thoralf and Paschereit, Christian Oliver},
    title = {Experimental and Numerical Advancement of the MGT Combustor Towards Higher Hydrogen Capabilities},
    volume = {Volume 3B: Combustion, Fuels, and Emissions},
    series = {Turbo Expo: Power for Land, Sea, and Air},
    year = {2022},
    doi = {10.1115/GT2022-82110},
}

@article{noble:2021,
    author = {Noble, David and Wu, David and Emerson, Benjamin and Sheppard, Scott and Lieuwen, Tim and Angello, Leonard},
    title = "{Assessment of Current Capabilities and Near-Term Availability of Hydrogen-Fired Gas Turbines Considering a Low-Carbon Future}",
    journal = {Journal of Engineering for Gas Turbines and Power},
    volume = {143},
    number = {4},
    year = {2021},
    month = {02},
    issn = {0742-4795},
    doi = {10.1115/1.4049346},
}

@article{jaeschke:2023,
    author = {Jaeschke, Alexander and Ćosić, Bernhard and Wassmer, Dominik and Paschereit, Christian Oliver},
    title = "{Experimental Investigation of a Multi Tube Burner for Premixed Hydrogen and Natural Gas Low Emission Combustion}",
    journal = {Journal of Engineering for Gas Turbines and Power},
    volume = {145},
    number = {12},
    year = {2023},
    month = {10},
    issn = {0742-4795},
    doi = {10.1115/1.4063378},
}

@article{hermeth:2024,
    author = {Hermeth, Sebastian and Panek, Lukasz and Witzel, Benjamin and Grandt, Christopher and Koestlin, Berthold and Wanjura, Stefan and Teuber, Hannes and et al.},
    title = "{100\% H2 DLE Gas Turbine Combustion Technology Platform Development}",
    volume = {Volume 3B: Combustion, Fuels, and Emissions},
    journal = {Proceedings of ASME Turbo Expo 2024},
    series = {Turbo Expo: Power for Land, Sea, and Air},
    year = {2024},
    month = {06},
    doi = {10.1115/GT2024-128517},
}

@article{zhou:2024,
title = {Hydrogen-fueled gas turbines in future energy system},
journal = {International Journal of Hydrogen Energy},
volume = {64},
year = {2024},
issn = {0360-3199},
doi = {https://doi.org/10.1016/j.ijhydene.2024.03.327},
author = {Haiqin Zhou and Jiye Xue and Haobu Gao and Nan Ma}
}

@article{jaeschke:2025,
    author = {Jaeschke, Alexander and Wassmer, Dominik and Paschereit, Christian Oliver},
    title = {Influence of Variable Fuel Staging on the Flow Structures in a Multi-Jet Burner Operated on Lean-Premixed Hydrogen},
    volume = {Volume 3B: Combustion, Fuels and Emissions},
    journal = {Proceedings Turbo Expo},
    series = {Turbo Expo},
    year = {2025},
    month = {06},
    doi = {10.1115/GT2025-153254}
}

@article{ax:2025,
    author = {Ax, Holger and Lammel, Oliver and Lückerath, Rainer and Gray, Joshua and Witzel, Benjamin and Blätte, Lutz and Köstlin, Berthold},
    title = {Investigation of Critical Operating Conditions for Hydrogen Flames Under Typical Gas Turbine Conditions},
    volume = {Volume 3A: Combustion, Fuels and Emissions},
    journal = {Proceedings Turbo Expo},
    series = {Turbo Expo},
    year = {2025},
    month = {06},
    doi = {10.1115/GT2025-152427}
}

@article{kalantari:2015,
    author = {Kalantari, Alireza and Sullivan-Lewis, Elliot and McDonell, Vincent},
    title = "{Flashback Propensity of Turbulent Hydrogen–Air Jet Flames at Gas Turbine Premixer Conditions}",
    journal = {Journal of Engineering for Gas Turbines and Power},
    volume = {138},
    number = {6},
    year = {2015},
    month = {11},
    issn = {0742-4795},
    doi = {10.1115/1.4031761},
}

@article{Eichler:2011,
title	= {Experimental Investigation of Turbulent Boundary Layer Flashback Limits for Premixed Hydrogen-Air Flames Confined in Ducts},
author	= {Eichler, Christian and Baumgartner, Georg and Sattelmayer, Thomas},
journal	= {J. Eng. Gas Turb. Power},
year	= {2011},
doi = {10.1115/1.4004149},
}

@article{Elbe:1945,
title	= {Further Studies of the Structure and Stability of Burner Flames},
author	= {von Elbe, Guenther and Mentser, Morris},
journal	= {J. Chemical Phys.},
year	= {1945},
}

@article{porath:2025,
title = {Low velocity streaks combined with intrinsic flame instabilities provoke boundary layer flashback in a turbulent premixed jet-stabilized hydrogen flame},
journal = {Combustion and Flame},
volume = {278},
year = {2025},
issn = {0010-2180},
doi = {https://doi.org/10.1016/j.combustflame.2025.114236},
author = {P. Porath and L.A. Altenburg and S.A. Klein and M.J. Tummers and A. Ghani}
}

@article{park:2025,
title = {Boundary layer flashback limits and flame dynamics of turbulent premixed hydrogen-air flames},
journal = {Combustion and Flame},
volume = {273},
year = {2025},
issn = {0010-2180},
doi = {https://doi.org/10.1016/j.combustflame.2024.113923},
author = {Jaehyun Park and Kyu Tae Kim}
}

@article{gruber2021,
title = {Direct Numerical Simulation of hydrogen combustion at auto-ignitive conditions: Ignition, stability and turbulent reaction-front velocity},
journal = {Combustion and Flame},
volume = {229},
pages = {111385},
year = {2021},
issn = {0010-2180},
doi = {https://doi.org/10.1016/j.combustflame.2021.02.031},
author = {Andrea Gruber and Mirko R. Bothien and Andrea Ciani and Konduri Aditya and Jacqueline H. Chen and Forman A. Williams},
}

@article{berger:2022,
title = {Synergistic interactions of thermodiffusive instabilities and turbulence in lean hydrogen flames},
journal = {Combustion and Flame},
volume = {244},
year = {2022},
issn = {0010-2180},
doi = {https://doi.org/10.1016/j.combustflame.2022.112254},
author = {Lukas Berger and Antonio Attili and Heinz Pitsch}
}

@article{Duan:2014,
title	= {Influence of Burner Material, Tip Temperature, and Geometrical Flame Configuration on Flashback Propensity of H2-Air Jet Flames},
author	= {Duan, Zhixuan and Shaffer, Brendan and McDonell, Vincent and Baumgartner, Georg and Sattelmayer, Thomas},
journal	= {J. Eng. Gas Turb. Power},
year	= {2014},
doi = {10.1115/1.4025359},
}

@article{Baumgartner:2015,
    author = {Baumgartner, Georg and Boeck, Lorenz R. and Sattelmayer, Thomas},
    title = "{Experimental Investigation of the Transition Mechanism From Stable Flame to Flashback in a Generic Premixed Combustion System With High-Speed Micro-Particle Image Velocimetry and Micro-PLIF Combined With Chemiluminescence Imaging}",
    journal = {Journal of Engineering for Gas Turbines and Power},
    volume = {138},
    number = {2},
    year = {2015},
    month = {09},
    issn = {0742-4795},
    doi = {10.1115/1.4031227},
}

@article{kalantari:2017,
title = {Boundary layer flashback of non-swirling premixed flames: Mechanisms, fundamental research, and recent advances},
journal = {Progress in Energy and Combustion Science},
volume = {61},
year = {2017},
issn = {0360-1285},
doi = {https://doi.org/10.1016/j.pecs.2017.03.001},
author = {Alireza Kalantari and Vincent McDonell}
}

@article{hoferichter2017,
author = {Vera Hoferichter and Christoph Hirsch and Thomas Sattelmayer},
title = {Analytic prediction of unconfined boundary layer flashback limits in premixed hydrogen–air flames},
journal = {Combustion Theory and Modelling},
volume = {21},
number = {3},
year = {2017},
publisher = {Taylor and Francis},
doi = {10.1080/13647830.2016.1240832}
}

@article{zurnedden:2025,
    author = {zur Nedden, Philipp Maximilian and Montagne, Oscar Luis and Peisdersky, Christoph and Eck, Mattias Ettore Giulio and Lückoff, Finn Simon and Orchini, Alessandro and Paschereit, Christian Oliver},
    title = {Investigation of Flashback Limits and Detection Strategies for Jet-Stabilized Hydrogen Flames},
    volume = {Volume 3A: Combustion, Fuels and Emissions},
    journal = {Proceedings Turbo Expo},
    series = {Turbo Expo},
    year = {2025},
    month = {06},
    doi = {10.1115/GT2025-152906}
}

@article{beuth2023, 
    author = {Beuth, Jan Paul and Reumschüssel, Johann Moritz and von Saldern, Jakob G. R. and Wassmer, Dominik and Ćosić, Bernhard and Paschereit, Christian Oliver and Oberleithner, Kilian},
    title = "{Thermoacoustic Characterization of a Premixed Multi Jet Burner for Hydrogen and Natural Gas Combustion}",
    journal = {Journal of Engineering for Gas Turbines and Power},
    volume = {146},
    number = {4},
    year = {2023},
    month = {12},
    doi = {10.1115/1.4063692},
}

@article{kang:2021,
title = {Combustion dynamics of multi-element lean-premixed hydrogen-air flame ensemble},
journal = {Combustion and Flame},
volume = {233},
year = {2021},
issn = {0010-2180},
doi = {https://doi.org/10.1016/j.combustflame.2021.111585},
author = {Hyebin Kang and Kyu Tae Kim}
}

@article{jin:2021,
title = {Experimental investigation of combustion dynamics and NOx/CO emissions from densely distributed lean-premixed multinozzle CH4/C3H8/H2/air flames},
journal = {Combustion and Flame},
volume = {229},
year = {2021},
issn = {0010-2180},
doi = {https://doi.org/10.1016/j.combustflame.2021.111410},
author = {Ukhwa Jin and Kyu Tae Kim},
}

@misc{ge:2025,
  author = {GE Vernova press release},
  title = {GE Vernova validates 100 percent hydrogen-fueled DLN combustor technology aiming to decarbonize its industrial B- and E-Class gas turbines, GE Vernova News},
  year = {2025},
  howpublished = {\url{https://www.gevernova.com/news/press-releases/ge-vernova-validates-100-hydrogen-fueled-}\newline\url{dln-combustor-technology-aiming-6}},
  urldate = {2025-11-27}
}

@article{vonsaldern2025,
title = {Low-frequency streaky structures in turbulent hydrogen jet flames},
journal = {Combustion and Flame},
volume = {278},
year = {2025},
issn = {0010-2180},
doi = {https://doi.org/10.1016/j.combustflame.2025.114231},
author = {Jakob G.R. {von Saldern} and Jan Paul Beuth and Johann Moritz Reumschüssel and Alexander Jaeschke and Christian Oliver Paschereit and Kilian Oberleithner}
}

@book{gibson2021,
  title={Additive Manufacturing Technologies},
  author={Gibson, I. and Rosen, D. and Stucker, B. and Khorasani, M.},
  isbn={9783030561260},
  DOI={https://doi.org/10.1007/978-3-030-56127-7}, 
  series={Engineering},
  year={2021},
  publisher={Springer Cham}
}

@article{cantera2018goodwin,
  title={Cantera: An object-oriented software toolkit for chemical kinetics, thermodynamics, and transport processes},
  author={Goodwin, David G and Speth, Raymond L and Moffat, Harry K and Weber, Bryan W},
  journal={Zenodo},
  year={2018}
}

@article{gri3,
  title={GRI-Mech 3.0},
  author={Gregory P. Smith and David M. Golden and Michael Frenklach and Nigel W. Moriarty and Boris Eiteneer and Mikhail Goldenberg and C. Thomas Bowman and et al.},
  url = { http://www.me.berkeley.edu/gri_mech/}
}

@article{schiavone:2024,
title = {On the adequacy of OH* as heat release marker for hydrogen–air flames},
journal = {Proceedings of the Combustion Institute},
volume = {40},
number = {1},
year = {2024},
issn = {1540-7489},
doi = {https://doi.org/10.1016/j.proci.2024.105248},
author = {Francesco G. Schiavone and Andrea Aniello and Eleonore Riber and Thierry Schuller and Davide Laera}
}

@article{schmidt2020, 
title={Instability of forced planar liquid jets: mean field analysis and nonlinear simulation}, 
volume={883}, 
DOI={10.1017/jfm.2019.855}, 
journal={Journal of Fluid Mechanics}, 
author={Schmidt, S. and Oberleithner, K.}, 
year={2020}, 
pages={A7}
}

@article{fellouah:2009,
title = {Reynolds number effects within the development region of a turbulent round free jet},
journal = {International Journal of Heat and Mass Transfer},
volume = {52},
number = {17},
year = {2009},
issn = {0017-9310},
doi = {https://doi.org/10.1016/j.ijheatmasstransfer.2009.03.029},
author = {H. Fellouah and C.G. Ball and A. Pollard}
}

@article{psomoglou:2024,
author = {Ianos Psomoglou and Burak Goktepe and Andrew Crayford and Phil Bowen and Steve Morris and Nick Jones},
title = {Influence of AM Generated Burner Surface Roughness on NOx Emissions and Operability of Hydrogen-Rich Fuels},
journal = {Combustion Science and Technology},
year = {2024},
publisher = {Taylor and Francis},
doi = {10.1080/00102202.2024.2390699}
}

@article{ding2017,
  title = {Numerical study of the influence of wall roughness on laminar boundary layer flashback},
  author = {Ding, Shuyu and Huang, Kai and Han, Yifan and Valiev, Damir},
  journal = {Phys. Rev. Fluids},
  volume = {6},
  issue = {2},
  year = {2021},
  month = {02},
  publisher = {American Physical Society},
  doi = {10.1103/PhysRevFluids.6.023201}
}

@article{vivoli:2025,
    author = {Vivoli, Robin and Pugh, Daniel and Goktepe, Burak and Hewlett, Sally and Giles, Anthony and Marsh, Richard and et al.},
    title = {Surface Roughness Effects on the Operability and Performance of a Hydrogen Jet Burner},
    journal = {Journal of Engineering for Gas Turbines and Power},
    volume = {148},
    year = {2025},
    month = {10},
    doi = {10.1115/1.4069474}
}

@article{kiymaz:2025,
title = {Numerical analysis of quenching distance in laminar premixed hydrogen and methane flames},
journal = {Fuel},
volume = {396},
year = {2025},
issn = {0016-2361},
doi = {https://doi.org/10.1016/j.fuel.2025.135111},
author = {Tahsin Berk Kıymaz and Nijso Beishuizen and Jeroen {van Oijen}}
}

@article{kosaka2018,
title = {Wall heat fluxes and CO formation/oxidation during laminar and turbulent side-wall quenching of methane and DME flames},
journal = {International Journal of Heat and Fluid Flow},
volume = {70},
year = {2018},
issn = {0142-727X},
doi = {https://doi.org/10.1016/j.ijheatfluidflow.2018.01.009},
author = {Hidemasa Kosaka and Florian Zentgraf and Arne Scholtissek and Lothar Bischoff and Thomas Häber and Christian Hasse and Andreas Dreizler and et al.}
}

@article{zirwes2021,
    author       = {Zirwes, Thorsten and Häber, Thomas and Zhang, Feichi and Kosaka, Hidemasa and Dreizler, Andreas and Steinhausen, Matthias and Hasse, Christian and et al.},
    year         = {2021},
    title        = {Numerical Study of Quenching Distances for Side-Wall Quenching Using Detailed Diffusion and Chemistry},
    volume       = {106},
    journal      = {Flow, turbulence and combustion},
    doi          = {10.1007/s10494-020-00215-0},
    publisher    = {{Springer}},
    issn         = {1386-6184, 1573-1987}
}

@article{kuwata:2020, 
title={Direct numerical simulation of turbulent conjugate heat transfer in a porous-walled duct flow}, 
volume={904}, 
DOI={10.1017/jfm.2020.669}, 
journal={Journal of Fluid Mechanics}, 
author={Kuwata, Y. and Tsuda, K. and Suga, K.}, 
year={2020}
}

@article{suga:2020, 
title={Characteristics of turbulent square duct flows over porous media}, 
volume={884}, 
DOI={10.1017/jfm.2019.914}, 
journal={Journal of Fluid Mechanics}, 
author={Suga, Kazuhiko and Okazaki, Yuki and Kuwata, Yusuke}, 
year={2020}
}

@article{fruzza:2024,
title = {The importance of Soret effect, preferential diffusion, and conjugate heat transfer for flashback limits of hydrogen-fueled perforated burners},
journal = {Proceedings of the Combustion Institute},
volume = {40},
year = {2024},
issn = {1540-7489},
doi = {https://doi.org/10.1016/j.proci.2024.105581},
author = {Filippo Fruzza and Hongchao Chu and Rachele Lamioni and Temistocle Grenga and Chiara Galletti and Heinz Pitsch}
}

@article{wang:2022,
title = {Linear instability of a premixed slot flame: Flame transfer function and resolvent analysis},
journal = {Combustion and Flame},
volume = {240},
year = {2022},
issn = {0010-2180},
doi = {https://doi.org/10.1016/j.combustflame.2022.112016},
author = {Chuhan Wang and Thomas L. Kaiser and Max Meindl and Kilian Oberleithner and Wolfgang Polifke and Lutz Lesshafft}
}

@article{garnaud:2013, 
title={The preferred mode of incompressible jets: linear frequency response analysis}, 
volume={716}, 
DOI={10.1017/jfm.2012.540}, 
journal={Journal of Fluid Mechanics}, 
author={Garnaud, X. and Lesshafft, L. and Schmid, P. J. and Huerre, P.}, 
year={2013}
}
\end{document}